\documentclass[a4paper]{jpconf}
\usepackage{graphicx}
\usepackage{amssymb,amsmath}

\newcommand{\ts}{\hspace{0.5pt}}
\newcommand{\nts}{\hspace{-0.5pt}}

\newcommand{\dd} {\,\mathrm{d}}

\newcommand{\ZZ}{\mathbb{Z}}
\newcommand{\RR}{\mathbb{R}\ts}
\newcommand{\NN}{\mathbb{N}}
\newcommand{\XX}{\mathbb{X}}
\newcommand{\vG}{\varGamma}
\newcommand{\cL}{\mathcal{L}}
\newcommand{\vL}{\varLambda}

\newcommand{\myfrac}[2]{\frac{\raisebox{-2pt}{$#1$}}{\raisebox{0.5pt}{$#2$}}}

\begin{document}
\title{Surprises in aperiodic diffraction}

\author{Michael Baake$^{1}$ and Uwe Grimm$^{2}$}

\address{$^{1}$ Fakult\"{a}t f\"{u}r Mathematik, Universit\"{a}t Bielefeld,
Postfach 100131, 33501 Bielefeld, Germany}
\address{$^{2}$ Department of Mathematics and Statistics, 
The Open University, Walton Hall, Milton Keynes MK7 6AA, UK}

\ead{mbaake@math.uni-bielefeld.de and u.g.grimm@open.ac.uk}

\begin{abstract}
  Mathematical diffraction theory is concerned with the diffraction
  image of a given structure and the corresponding inverse problem of
  structure determination. In recent years, the understanding of
  systems with continuous and mixed spectra has improved
  considerably. Moreover, the phenomenon of homometry shows various
  unexpected new facets. Here, we report on some of the recent results
  in an exemplary and informal fashion.
\end{abstract}

\section{Introduction}

The diffraction measure is a characteristic quantity of a translation
bounded measure $\omega$ on Euclidean space (or on any locally compact
Abelian group). It emerges as the Fourier transform
$\widehat{\gamma}$ of the autocorrelation measure $\gamma$ of $\omega$,
and has important applications in crystallography, because it
describes the outcome of kinematic diffraction (from X-rays or neutron
scattering, say). In recent years, initiated by the discovery of
quasicrystals (which are non-periodic but nevertheless show pure Bragg
diffraction), a systematic study by many people has produced a
reasonably satisfactory understanding of the class of measures with a
pure point diffraction measure, meaning that $\widehat{\gamma}$ is a
pure point measure, without any continuous component.

Clearly, reality is more complicated than that, in the sense that real
world structures will (and do) show lots of continuous components as
well. Unfortunately, the methods around dynamical systems that are
used to establish pure point spectra do not seem to extend to the
treatment of systems with mixed spectrum, at least not in sufficient
generality. Nevertheless, systems with continuous spectral components
have recently been investigated also from a rigorous mathematical
point of view, with a number of unexpected results. In particular, the
phenomenon of homometry becomes more subtle, as we will see below.

In this informal exposition, we summarise some classic results that
have recently resurfaced in the context of diffraction, with special
emphasis on singular continuous and absolutely continuous spectra. For
proofs (and the formal details) we refer to the original papers or to
work in progress.

\section{Mathematical setting and pure point spectra}

Below, we mainly consider various (weighted) Dirac combs on the real
line, with support on $\ZZ$ (the set of integers), and their
diffraction. Even in this simple and seemingly restricted setting,
unexpected phenomena show up.  If $\omega = \sum_{n\in\ZZ} w(n)
\delta_{n}$, with $\delta_{n}$ the normalised point measure at $n$,
the \emph{natural autocorrelation} (if it exists) of $\omega$ is
\begin{equation} \label{eq:auto-def}
   \gamma \, = \, \omega \circledast \widetilde{\omega} \,
    := \lim_{N\to\infty} \frac{\omega^{}_{N} \nts * 
        \widetilde{\omega^{}_{N}}}{2\ts N}\ts ,
\end{equation}
which we view as a measure on $\RR$. Here, $\omega^{}_{N} = \omega
|_{[-N,N]} $ is the restriction of $\omega$ to the closed interval
$[-N,N]$, and $\widetilde{\mu}\ts$ is the `flipped over' measure,
defined by $\widetilde{\mu} (g) = \overline{\mu (\widetilde{g}\ts )}$
with $g$ a continuous test function of compact support and
$\widetilde{g} (x) = \overline{g (-x) }$. Here, and in analogous
situations below, the bar denotes complex conjugation. We will only
consider situations where the limit in \eqref{eq:auto-def} exists,
either always (in the deterministic cases) or almost surely (in the
probabilistic cases). An explicit calculation shows that
$\mathrm{supp} (\omega) \subset \ZZ$ implies that $\gamma$ must be of
the form
\begin{equation} \label{eq:auto-form}
   \gamma \, = \sum_{m\in\ZZ} \eta(m) \, \delta_{m}
\end{equation}
with the autocorrelation coefficients 
\begin{equation} \label{eq:auto-coeff}
    \eta(m) \,  =  \lim_{N\to\infty} \frac{1}{2N \nts +1}
        \sum_{n=-N}^{N}  w(n)\ts \overline{w(n-m)} \ts .
\end{equation}
The existence of the limit in \eqref{eq:auto-def} is equivalent to the
existence of the $\eta(m)$ for all $m\in\ZZ$.

By construction, $\gamma$ is a positive definite measure, and hence
Fourier transformable. The result is $\widehat{\gamma}$, the
\emph{diffraction measure}, which is a positive, translation bounded
measure on $\RR$. It describes (kinematic) diffraction from the
measure $\omega$; compare \cite{Cowley} for background and
\cite{Hof,BM} for the development of this approach to diffraction
theory.  One of the benefits of this approach is the unique
decomposition
\begin{equation} \label{eq:decomp}
   \widehat{\gamma} = \bigl(\widehat{\gamma}\bigr)_{\mathsf{pp}}
   + \bigl(\widehat{\gamma}\bigr)_{\mathsf{sc}}      
   + \bigl(\widehat{\gamma}\bigr)_{\mathsf{ac}}
\end{equation}
of the diffraction measure into its pure point, singular continuous
and absolutely continuous parts, the latter splitting relative to Lebesgue
measure $\lambda$, which is the natural reference measure for volume
in Euclidean space.

\smallskip
The best known example is the lattice Dirac comb $\omega =
\delta^{}_{\ZZ} = \sum_{n\in\ZZ} \delta_{n}$, with autocorrelation
$\gamma = \delta^{}_{\ZZ}$ and diffraction $\widehat{\gamma} =
\delta^{}_{\ZZ}$. Here, $\gamma$ follows from an elementary
calculation, while the formula for $\widehat{\gamma}$ is a consequence
of the \emph{Poisson summation formula}
\begin{equation} \label{eq:PSF}
   \widehat{\delta^{}_{\nts\vG}} \, = \,
   \mathrm{dens} \ts (\vG) \, \delta^{}_{\nts \vG^{*}}
\end{equation}
for an arbitrary lattice $\vG$ (in $\RR^{d}$) of density
$\mathrm{dens} \ts (\vG)$, with dual lattice $\vG^{*} $, the latter
defined by $\vG^{*} := \{ x\in\RR^{d} \mid xy \in \ZZ \mbox{ for all }
y\in\vG\}$; see \cite{BM} and references therein. The integer lattice
is perhaps the simplest example for a system with pure point
diffraction. More generally, if $\omega$ is any $\ZZ$-periodic
measure, it can be written as $\omega = \mu * \delta^{}_{\ZZ}$, with
$\mu$ a finite measure. This leads to $\gamma = (\mu *
\widetilde{\mu}\ts ) * \delta^{}_{\ZZ}$ and $\widehat{\gamma} = \lvert
\widehat{\mu} \rvert^{2} \, \delta^{}_{\ZZ}$, where $\lvert
\widehat{\mu} \rvert^{2}$ is a continuous positive function on
$\RR$. Its values at integer points are the \emph{intensities} of the
Bragg peaks.

A similarly nice formula formula holds for regular model sets.  Within
a given cut and project scheme (CPS) $(\RR^{d},\RR^{m},\cL)$ with
lattice $\cL \subset \RR^{d+m}$, consider $\vL = \{ x\in L \mid x^{\star} \in
W\}$, where $L=\pi (\cL)$ is the projection of $\cL$ in $\RR^{d}$, $W$
is the (compact) window in $\RR^{m}$ (with boundary of measure $0$),
and $\star$ denotes the star map of the CPS. By the general model set
theorem \cite{Hof,M,BM}, the diffraction $\widehat{\gamma}$ of the
Dirac comb $\delta^{}_{\nts \vL}$ is then the pure point measure
\begin{equation} \label{eq:model-set}
  \widehat{\gamma}\, = \sum_{k\in L^{\circledast}}
  \lvert A(k) \rvert^{2}\, \delta_{k} \ts ,
\end{equation}
where $L^{\circledast}=\pi (\cL^{*})$ is the projection of the
dual lattice $\cL^{*}$ and the amplitude $A(k)$ is given by
\[
   A(k)\, = \, \frac{\mathrm{dens} (\vL)}{\mathrm{vol} (W)}
   \int_{W} e^{2 \pi i k^{\star} y} \dd y \, = \,
   \frac{\mathrm{dens} (\vL)}{\mathrm{vol} (W)}
   \, \widehat{1^{}_{W}} (-k^{\star})\ts .
\]
This formula has various generalisations, for instance to model sets
with other internal spaces or to weighted model sets; see
\cite{M,BM} and references therein for more.

\smallskip

As an example, we consider the \emph{period doubling sequence}, as
defined by the substitution
\begin{equation} \label{eq:pd-def}
   \varrho = \varrho^{}_{\mathrm{pd}}\! : \, 
   a \mapsto ab \, , \; b \mapsto aa \ts .
\end{equation}    
A two-sided sequence can be obtained from a fixed point of
$\varrho^{2}$ via the iteration
\begin{equation} \label{eq:pd-iter}
    a | a \xrightarrow{\,\varrho^{2}\,} abaa | abaa 
    \xrightarrow{\,\varrho^{2}\,}
    \ldots \longrightarrow w = \varrho^{2} (w) \ts ,
\end{equation}
with convergence in the (obvious) product topology. Here and below, we
write two-sided sequences as $w = \ldots w^{}_{-2} w^{}_{-1} |
w^{}_{0} w^{}_{1} \ldots$ and use $|$ to mark the origin. Note
that $a|a$ in \eqref{eq:pd-iter} is a legal seed, so that $w$ is a
fixed point of $\varrho^{2}$ in the strict sense.

We attach a Dirac comb to $w$ by $\omega = \sum_{n\in\ZZ} h(w_{n} )\,
\delta_{n}$, with $h(a) = h_{+}$ and $h(b) = h_{-}$. This turns out to
define a (weighted) regular model set \cite{BMS,BM}, so that the
diffraction is indeed a pure point measure. It is given by
\[
     \widehat{\gamma^{}_{\mathrm{pd}}}\, =  
     \sum_{k\in L^{\circledast}}
     \bigl| h_{+} \ts A(k) + h_{-} \ts B(k) \bigr|^{2} 
     \, \delta^{}_{k} \ts ,
\]
where $L^{\circledast} = \bigcup_{\ell\ge 1} \ZZ/2^{\ell}
= \bigl\{ \frac{m}{2^r} \mid (r=0 \ts , \, m\in\ZZ)
\mbox{ or } (r\ge 1 \ts , \, m \mbox{ odd}) \bigr\}$ is the
Fourier module of the period doubling sequence. The
amplitudes read
\[
   A(k) = \myfrac{2}{3\cdot (-2)^{r}}\ts
   e^{2\pi i k}   \quad \mbox{ and } \quad
   B(k) = \delta^{}_{r,0} - A(k)\ts ,
\]
where we implicitly refer to the parametrisation of $L^{\circledast}$.
We state this formula here without further details; see \cite{BMS,BM}
for a proof.

\smallskip

Let us briefly mention the \emph{homometry} problem. It refers to the
possibility that distinct measures can still possess the same
autocorrelation. An interesting example is constructed in \cite{GM},
based on $6$-periodic Dirac combs of the form $\delta^{}_{6\ZZ} *
\sum_{j=0}^{5} c^{}_{j} \ts \delta^{}_{j}$.  In particular, the two
choices of Table~\ref{tab:gm} lead to the same autocorrelation -- and,
in fact, even to identical correlation functions up to order $5$. It
is only the order $6$ correlation that tells the two combs apart.

\begin{table}
\caption{\label{tab:gm}Weights for the homometric pair from \cite{GM}.}
\begin{center}
\begin{tabular}{ccccccc}
\br
$j$     & 0 & 1 & 2 & 3 & 4 & 5 \\
\mr
$c_{j}$ & 11 & 25 & 42 & 45 & 31 & 14 \\
$c_{j}$ & 10 & 21 & 39 & 46 & 35 & 17 \\
\br
\end{tabular}
\end{center}
\end{table}

For model sets, there are further possibilities, because different
windows can have the same covariogram. The latter (for a compact set
$K$) is defined as $\mathrm{cvg}^{}_{K} (x) = \mathrm{vol} \bigl(K\cap
(x+K)\bigr) = \bigl(1^{}_{K} \nts * \widetilde{1^{}_{K}}\bigr) (x)$.
This phenomenon results in identical autocorrelations and hence in
homometric model sets. A simple planar example was constructed in
\cite{BG}; see also \cite{GB,DM1,DM2}.  We skip further details here
and shift our attention to continuous spectra now.

\section{Singular continuous spectra}

The paradigm of a singular continuous measure is the distribution
function for the classic middle-thirds Cantor measure, which is also
called the Devil's staircase. Here, we show a class of structures that
lead to somewhat similar functions, yet with significant differences.

\smallskip

The classic Thue-Morse (or Prouhet-Thue-Morse) sequence \cite{A} can be
defined via a fixed point of the substitution
\begin{equation} \label{eq:tm-def}
   \varrho = \varrho^{}_{\mathrm{TM}} \! : \;
   1 \mapsto 1 \bar{1} \, , \; \bar{1} \mapsto \bar{1} 1
\end{equation}
on the binary alphabet $\{ 1,\bar{1}\}$. The one-sided fixed point
starting with $1$ reads
\begin{equation}   \label{eq:tm-fix}
    v = v^{}_{0} v^{}_{1} v^{}_{2} \ldots = 
    1 \bar{1} \bar{1} 1 \bar{1} 1 1 \bar{1} \ldots \ts ,
\end{equation}
while $\bar{v}$ is the fixed point starting with $\bar{1}$. One can
now define a two-sided sequence $w$ by
\[
    w_i  =  \begin{cases}   v_i ,  & i\ge 0 , \\
                v_{-i-1}, & i<0 .  \end{cases}
\]
It is easy to check that $w$ defines a $2$-cycle under $\varrho$, and
hence a fixed point for $\varrho^{2}$. Since the central seed
$1|1$ is legal, an iteration of $\varrho^{2}$ applied to it
converges to $w$ in the product topology.

The sequence $w$ defines a dynamical system (under the action of the
group $\ZZ$). Its compact space is the (discrete) \emph{hull},
obtained as the closure of the $\ZZ$-orbit of $w$,
\[
      \XX^{}_{\mathrm{TM}} = \overline{ \{ S^{i} w \mid i\in\ZZ\} }\ts ,
\]
where $S$ denotes the shift operator and the closure is taken in the
local (or product) topology, where two sequences are close when they
agree on a large segment around the origin. Now,
$(\XX^{}_{\mathrm{TM}}, \ZZ)$ is a strictly ergodic dynamical system
(hence uniquely ergodic and minimal \cite{Q,W}). Its unique invariant
probability measure is given via the (absolute) frequencies of finite
words (or patches) as the measures of the corresponding cylinder sets,
which then generate the $\sigma$-algebra.

Here, we are interested in the diffraction of the (signed) Dirac comb
$ \omega^{}_{\mathrm{TM}} = \sum_{n\in\ZZ} w_n \ts \delta_n$, where we
interpret $1$ and $\bar{1}$ as weights $1$ and $-1$. Its
autocorrelation exists, as a consequence of unique ergodicity, and is
actually the same measure for \emph{all} sequences of
$\XX^{}_{\mathrm{TM}}$.  We can now employ the special structure of
our fixed point $w$ to analyse it. The autocorrelation is of the
general form \eqref{eq:auto-form}, with coefficients $\eta (0) = 1$,
$\eta (-m) = \eta (m)$ for all $m\in\ZZ$ and
\begin{equation} \label{eq:tm-coeff}
     \eta (m) \,  = \lim_{N\to\infty} \frac{1}{N}
        \sum_{n=0}^{N-1} v_{n} v_{n+m}
\end{equation}
for all $m\ge 0$. Here, the structure of $w$ and its relation to $v$
was used to derive \eqref{eq:tm-coeff} from \eqref{eq:auto-coeff}.
Observing that $v$ satisfies $ v^{}_{2n} = v^{}_{n}$ and $v^{}_{2n+1}
= \bar{v}^{}_{n}$ for all $n\ge 0$, one can derive from
\eqref{eq:tm-coeff} the recursions
\begin{equation} \label{eq:tm-rec}
     \eta (2m) = \eta (m) \quad \mbox{and} \quad
     \eta(2m+1) = - \myfrac{1}{2} \bigl( \eta (m) + \eta (m+1) \bigr) ,
\end{equation}
which actually hold for all $m\in\ZZ$. One finds $\eta (\pm 1) =
-1/3$ from solving the recursion for $m=0$ and $m=-1$ with
$\eta(0)=1$, while all other values are then recursively determined.

To analyse the diffraction measure $\widehat{\gamma}$ of the TM
sequence (following \cite{Mah,Kaku}), one can start with its pure
point part. Defining $\varSigma (N) = \sum_{n=-N}^{N} \bigl( \eta (n)
\bigr)^{2}$, one derives via the recursion \eqref{eq:tm-rec} that
$\varSigma (4N) \le \frac{3}{2}\varSigma(2N)$, which implies $\frac{1}{N}
\varSigma (N) \xrightarrow{\, N\to\infty \,} 0$. By Wiener's
criterion \cite{Wie}, this means $\bigl( \widehat{\gamma}
\bigr)_{\mathsf{pp}}=0$, so that $\widehat{\gamma}$ is a continuous
measure.

\begin{figure}
\begin{center}
\includegraphics[width=0.8\textwidth]{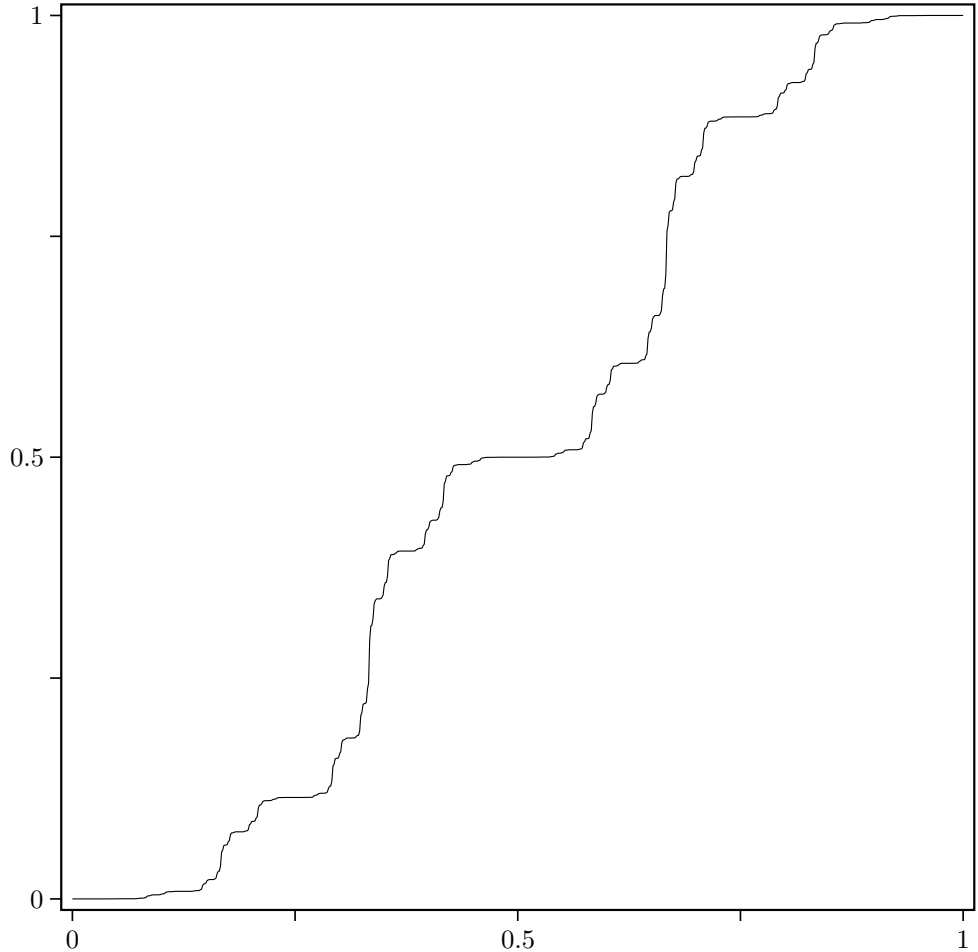}
\end{center}
\caption{\label{fig1}The distribution function of the TM measure on $[0,1]$.}
\end{figure}

Defining the (continuous) distribution function $F$ via $F(x) =
\widehat{\gamma} \bigl( [0,x]\bigr)$, another consequence of
\eqref{eq:tm-rec} is the pair of functional relations
\[
   \dd F \bigl( \tfrac{x}{2} \bigr) \pm
   \dd F \bigl(\tfrac{x+1}{2} \bigr) = \biggl\{
   \begin{array}{c} 1 \\\!\!\! - \cos (\pi x)\!\!\! \end{array}
   \biggr\}   \dd F (x) \ts .
\]
Splitting $F$ into its \textsf{sc} and \textsf{ac} parts (which is
unique) now implies backwards that the recursion \eqref{eq:tm-rec}
holds separately for the two sets of autocorrelation coefficients,
$\eta^{}_{\mathsf{sc}}$ and $\eta^{}_{\mathsf{ac}}$, with yet unknown
initial conditions at $0$. An application of the Riemann-Lebesgue
lemma, however, forces $\eta^{}_{\mathsf{ac}} (0) =0$, and hence
$\eta^{}_{\mathsf{ac}} (m) = 0$ for all $m\in\ZZ$, so that also
$\bigl( \widehat{\gamma} \bigr)_{\mathsf{ac}}=0$; compare
\cite{Kaku}. This shows that $\widehat{\gamma}$, which is not the zero
measure, is purely singular continuous. Figure~\ref{fig1} shows an
image, where we have used the Volterra iteration
\[
     F^{}_{n+1} (x) = \myfrac{1}{2} \int_{0}^{2x}
     \bigl( 1 - \cos (\pi y)\bigr) F^{\ts \prime}_{n} (y)
     \dd y  \qquad \mbox{with} \qquad F^{}_{0} (x) = x
\]
to calculate $F$ with sufficient precision (note that $F(x+1) = F(x) +
1$, so that a display on $[0,1]$ suffices). In contrast to the Devil's
staircase, the TM function is \emph{strictly} increasing, which means
that there is no plateau (which would indicate a gap in the support
of $\widehat{\gamma}$); see \cite{BG08} and references therein for
details.

\smallskip
Despite the above result, the TM sequence is closely related to
the period doubling sequence, via the (continuous) block map
\begin{equation} \label{eq:block-map}
  \varphi \! : \quad 1\bar{1} \ts , \ts \bar{1}1 \mapsto a
  \, , \quad 11 \ts , \ts \bar{1}\bar{1} \mapsto b \, ,
\end{equation}
which defines an exact 2-to-1 surjection from the hull
$\XX^{}_{\mathrm{TM}}$ to $\XX^{}_{\mathrm{pd}}$.
\smallskip 

The TM sequence is often considered as a rare and special example,
which is misleading. To demonstrate the point (following \cite{Kea}),
consider the generalised Morse sequences defined by
\begin{equation} \label{eq:gen-tm-def}
   \varrho = \varrho^{}_{k,\ell} \! : \;
   1 \mapsto 1^{k} \bar{1}^{\ell} \, , \; 
   \bar{1} \mapsto \bar{1}^{k} 1^{\ell}
\end{equation}
for arbitrary $k,\ell\in\NN$. Here, the one-sided fixed point
starting with $v^{}_{0}=1$ satisfies
\begin{equation} \label{eq:gen-tm-seq}
  v^{}_{m(k+\ell)+r} = \begin{cases}
  v^{}_{m} , & \mbox{if } 0\le r < k \ts , \\
  \bar{v}^{}_{m} , & \mbox{if } k\le r < k+\ell \ts ,
  \end{cases}
\end{equation}
for $m\ge 0$.  A two-sided sequence can be constructed as above. Since
each choice of $k,\ell$ leads to a strictly ergodic dynamical system,
we know that all autocorrelation coefficients exist. Clearly (again
with $\bar{1}\, \widehat{=} -1$), we have $\eta(0)=1$, while several
possibilities exist to calculate $\eta(\pm 1) =
\frac{k+\ell-3}{k+\ell+1}$. In general, we obtain the recursion
\begin{equation} \label{eq:gen-tm-rec}
   \eta \bigl( (k+\ell) m + r\bigr) = \myfrac{1}{k+\ell}
   \bigl( \alpha^{}_{k,\ell,r}\, \eta(m) +
   \alpha^{}_{k,\ell,k+\ell-r} \, \eta(m+1) \bigr),
\end{equation}
with $\alpha^{}_{k,\ell,r} = k+\ell-r -2 \min (k,\ell,r,k+\ell-r)$,
which holds for all $m\in\ZZ$ and $0\le r\le k+\ell-1$.

The recursion can once again be used to show the absence of
pure point components (by Wiener's criterion) as well as that
of absolutely continuous components (by the Riemann-Lebesgue lemma),
thus establishing that each sequence in this family leads to a
signed Dirac comb with purely singular continuous diffraction.
The distribution function satisfies
\[
    F(x) = \widehat{\gamma} \bigl( [0,x] \bigr) =
    x + \sum_{m\ge 1} \frac{\eta(m)}{m\pi} \ts
    \sin (2\pi m x) \ts ,
\]
which is a uniformly converging series. Moreover, the measure
$\widehat{\gamma}$ has a (vaguely convergent) representation as an
infinite \emph{Riesz product}:\ With $\theta (x) :=
1+\frac{2}{k+\ell}\sum_{r=1}^{k+\ell-1} \alpha^{}_{k,\ell,r}\ts \cos
(2\pi r x)$, it is given by $ \prod_{n\ge 0} \theta \bigl(
(k+\ell)^{n} x \bigr)$. The entire analysis is thus completely
analogous to that of the original TM sequence and shows that the
latter is a typical example in an infinite family. Further details
(and proofs) will be given in \cite{BBGG}.

Let us finally mention that the block map \eqref{eq:block-map}
applies to any member of this family, and always gives a
2-to-1 cover of the hull that belongs to the generalised
period doubling substitution
\[
   \varrho^{\ts\prime} \! : \;
   a \mapsto b^{k-1} a b^{\ell-1} b \ts , \;
   b \mapsto b^{k-1} a b^{\ell-1} a \ts .
\]
Since we always have a coincidence in the sense of Dekking \cite{D},
they all define systems with pure point spectrum -- another
analogy to the classic case $k=\ell=1$.

\section{Absolutely continuous spectra}

The simplest example in this category is the Bernoulli (or coin
tossing) comb. Let $(W_{\nts n})_{n\in\ZZ}$ be a family of independent
and identically distributed (i.i.d.)~random variables. We assume that
$W$ represents this family and takes values $1$ and $-1$ with
probabilities $p$ and $1-p$. We then consider the random measure
\begin{equation} \label{eq:coin-def}
   \omega^{}_{\mathrm{B}} = \sum_{n\in\ZZ} 
         W_{\nts\nts n} \, \delta_{n} \ts ,
\end{equation}
which (almost surely in the probabilistic sense) has a natural
autocorrelation of the form \eqref{eq:auto-form}, with coefficients
$\eta^{}_{\mathrm{B}} (0) = 1$ and (a.s.) $\eta^{}_{\mathrm{B}} (m) =
(2p-1)^{2}$ for all $0\ne m\in\ZZ$, so that we have
\begin{equation} \label{eq:coin-auto}
    \gamma^{}_{\mathrm{B} }  =  (2p-1)^{2} \ts \delta^{}_{\ZZ}
    + 4\ts p(1-p)\ts \delta^{}_{0}
    \qquad \mbox{(a.s.).}
\end{equation}     
In particular, when $p=1/2$, it gives $\gamma^{}_{\mathrm{B} } =
\delta^{}_{0}$. Eq.~\eqref{eq:coin-auto} can be proved either by an
application of Birkhoff's ergodic theorem for $\ZZ$-action or by the
strong law of large numbers (SLLN). We refer to \cite{BG09} for
details and further references. Let us also mention that
\eqref{eq:coin-auto} can be rewritten as $\gamma^{}_{\mathrm{B} } =
\beta^{2} \ts \delta^{}_{\ZZ} + (1-\beta^{2})\ts \delta^{}_{0}$, where
$\beta$ has the meaning of the drift velocity of a one-dimensional
random walk on the line; compare \cite{Sh,BL}.

As a consequence, we obtain a diffraction measure of mixed type,
\begin{equation} \label{eq:coin-diffract}
    \widehat{\gamma^{}_{\mathrm{B}}} =  
    (2p-1)^{2} \ts \delta^{}_{\ZZ} + 
    4\ts p(1-p)\ts \lambda \qquad \mbox{(a.s.),}
\end{equation}
where $\lambda$ denotes Lebesgue measure. The special case of a fair
coin ($p=1/2$) leads to $\widehat{\gamma^{}_{\mathrm{B}}} =\lambda$,
which is purely absolutely continuous. More generally, when $W$ takes
the (possibly complex) values $h_{+}$ and $h_{-}$, each with
probability $1/2$, one obtains
\[
      \widehat{\gamma} = \big| \tfrac{h_{+} + h_{-}}{2} \big|^2 
      \ts\delta^{}_{\ZZ}
       + \big| \tfrac{h_{+} - h_{-}}{2} \big|^2 \ts\lambda  \ts ,
\]
which is another simple example of a mixed spectrum with pure point
and absolutely continuous components.
\smallskip 

An interesting deterministic counterpart is the \emph{Rudin-Shapiro
  sequence} \cite{R,S,Q,A}. Its quaternary version is usually defined
by the substitution
\[
   \varrho = \varrho^{}_{\mathrm{RS}} \! : \;
   a \mapsto ac \, , \; b \mapsto dc \, ,
   c \mapsto ab \, , \; d \mapsto db \ts .
\]
A fixed point $u=\varrho^{2} (u)$ can be constructed from the legal
seed $b|a$ via iteration of $\varrho^{2}$. The \emph{binary} version
follows then as the reduction $w=\varphi(u)$ with the mapping
$\varphi$ given by $\varphi(a)=\varphi(c)=1$ and
$\varphi(b)=\varphi(d)=-1$. The two-sided sequence $w$ satisfies
$w^{}_{-1} = -1$ and $w^{}_{0} = 1$ together with the recursion
\begin{equation} \label{eq:rs-rec}
   w(4n+\ell) = \begin{cases}
    w(n), & \mbox{for } \ell\in\{0,1\} , \\
  (-1)^{n+\ell} w(n), & \mbox{for } \ell\in\{2,3\} ,\
   \end{cases}
\end{equation}
for all $n\in\ZZ$. The sequence $w$ and its hull once again define a
strictly ergodic dynamical system, so that the natural autocorrelation
$\gamma^{}_{\mathrm{RS}}$ exists and is the same for all members of
the hull. It is of the general form \eqref{eq:auto-form}. To 
calculate it, one needs a little trick. Define the coefficients
\begin{equation} \label{eq:rs-coeff}
   \begin{array}{c} \eta(m)\!\! \\ \vartheta(m)\!\! \end{array}
   \biggr\}\, := \lim_{N\to\infty}\, \frac{1}{2N+1}
   \sum_{n=-N}^{N} w(n)\ts w(n+m)\, \biggl\{
   \begin{array}{c} \!\! 1 \\ \!\! (-1)^n \end{array}
\end{equation}
for $m\in\ZZ$. Note that all these limits exist, by Birkhoff's
ergodic theorem (this is immediately clear for $\eta(m)$, but
also follows for $\vartheta(m)$ on the basis of the quarternary
sequence $u$, which also gives rise to a strictly ergodic system);
compare \cite{PF} for an alternative approach.

It is easy to check that $\eta(0)=1$ and $\vartheta(0)=0$, while
a slightly more involved calculation reveals the following
coupled system of recursions,
\[
   \begin{split}
   \eta\ts(4m) &\,=\, \myfrac{1+(-1)^m}{2}\,\eta\ts(m)\ts , \qquad
   \eta\ts(4m\!+\!2) \,=\, 0\vphantom{\myfrac{(1-(1-)^m}{4}}\ts ,\\
   \eta\ts(4m\!+\!1) &\,=\, \myfrac{1-(-1)^m}{4}\,\eta\ts(m) + 
   \myfrac{(-1)^m}{4}\,\vartheta(m) - \myfrac{1}{4}\,\vartheta(m\!+\!1)\ts ,\\
   \eta\ts(4m\!+\!3) &\,=\, \myfrac{1+(-1)^m}{4}\,\eta\ts(m\!+\!1) - 
   \myfrac{(-1)^m}{4}\,\vartheta(m) + \myfrac{1}{4}\vartheta(m\!+\!1)\ts ,
   \end{split}
\]
together with
\[
   \begin{split}
   \vartheta(4m) &\,=\, 0\vphantom{\myfrac{1}{4}}\ts , \qquad
   \vartheta(4m\!+\!2) \,=\, \myfrac{(-1)^m}{2}\,\vartheta(m)+\myfrac{1}{2}\,
   \vartheta(m\!+\!1)\ts , \\ 
   \vartheta(4m\!+\!1) &\,=\, \myfrac{1-(-1)^m}{4}\,\eta\ts(m) - 
   \myfrac{(-1)^m}{4}\,\vartheta(m) +\myfrac{1}{4}\,\vartheta(m\!+\!1)\ts ,\\ 
   \vartheta(4m\!+\!3) &\,=\, -\myfrac{1+(-1)^m}{4}\,\eta\ts(m\!+\!1) -
   \myfrac{(-1)^m}{4}\,\vartheta(m) + \myfrac{1}{4}\,\vartheta(m\!+\!1)\ts .
   \end{split}
\]
These equations are valid for all $m\in\ZZ$. Extracting $\vartheta(1)=
\vartheta(-1)=0$ from the last two equations (with $m=0$ resp.\
$m=-1$), one can then recursively conclude that $\vartheta(m)=0$
for all $m\in\ZZ$ and $\eta(m)=0$ for all $m\ne 0$. This shows
\begin{equation} \label{eq:rs-diffract}
   \gamma^{}_{\mathrm{RS}} = \delta^{}_{0}
   \qquad \mbox{and} \qquad
   \widehat{\gamma^{}_{\mathrm{RS}}} = \lambda \ts .
\end{equation}
In particular, the (deterministic) diffraction \eqref{eq:rs-diffract}
of the binary RS sequence is the same as the (almost sure)
diffraction of the coin tossing sequence \eqref{eq:coin-diffract} for
$p=1/2$, despite the fact that their entropies are different ($0$
versus $\log (2)$); see \cite{HB,BG09} for further details.

\smallskip 

The situation is actually `worse' in the following sense. Let
$S\in\{\pm 1\}^{\ZZ}$ be a two-sided sequence, assumed ergodic, with
associated Dirac comb $\omega^{}_{S} = \sum_{n\in\ZZ} S_{n}\ts
\delta_{n}$ and autocorrelation $\gamma^{}_{S}$. The latter exists due
to the ergodicity assumption. Let $(W_{\nts\nts n})_{n\in\ZZ}$ be the
above family of i.i.d.\ random variables with values in $\{ \pm 1
\}$. Consider now the new random Dirac comb
\begin{equation} \label{eq:bern-def}
   \omega \, = \sum_{n\in\ZZ} S_{n}\ts W_{\nts \nts n}\, \delta_{n}\ts ,
\end{equation}
which we call the \emph{Bernoullisation} of the original sequence.  It
can be considered as a `randomisation via second thoughts', since the
action of the $W_{n}$ is to either keep $S_{n}$ (with probability $p$)
or to change its sign (with probability $1-p$). By another (slightly
more complicated) application of the SLLN, one finds \cite{BG09}
\begin{equation} \label{eq:bern-auto}
   \gamma = (2p-1)^{2} \ts \gamma^{}_{S} + 4\ts p (1-p) \ts \delta^{}_{0}
   \qquad \mbox{(a.s.)}
\end{equation}
together with the corresponding almost sure diffraction
$\widehat{\gamma} = (2p-1)^{2} \, \widehat{\gamma^{}_{S}}
+ 4\ts  p (1-p)\ts \lambda$. If $S$ is the binary RS sequence from
above, one finds $\gamma = \delta^{}_{0}$ and $\widehat{\gamma}
=\lambda$, \emph{independently} of the parameter $p$. This proves
that the entire family defined by \eqref{eq:bern-def} for the
RS sequence is homometric, with the entropy varying continuously
between $0$ and $\log(2)$; compare \cite{BLR} for the connection
between diffraction and entropy in the pure point case.

\section{Outlook}   

The diffraction measure is a useful tool, both for the understanding
of experiments and for various theoretical questions. Beyond the
well-studied case of pure point spectra, also continuous spectra are
explicitly accessible and sometimes (as above) even computable.
However, as was to be expected, various aspects are more involved, and
this includes the inverse problem. The latter is increasing in
complexity also with growing dimension. For instance, the Ledrappier
system \cite{L} in the plane has the same autocorrelation as the
Bernoulli comb (both living on $\ZZ^{2}$), but has rank-$1$ entropy
despite being genuinely two-dimensional.

Various generalisations exist to non-lattice systems, which are
best formulated via the theory of point processes \cite{BBM}.
Here, the classic Poisson process (of mean point density $1$)
is another example (with random positions) that gives diffraction
measure $\lambda$ when the points are randomly weighted with
$1$ and $-1$. It is clear that higher order correlations can
tell these systems apart, but many questions are still open.

Finally, more realistic models of randomness have to include
interactions. The theory of Gibbs measures is a necessary and useful
tool here, but only first steps in this direction have been taken; see
\cite{BBM} and references therein. Our present day understanding is
still rather limited.

\ack It is a pleasure to thank Shelomo I.\ Ben-Abraham, Daniel Lenz,
Robert V.\ Moody and Tom Ward for various discussions. This work was
supported by the German Research Council (DFG), within the CRC 701, by
EPSRC, via Grant EP/D058465, and by a Leverhulme Trust Visiting
Professorship Grant.

\end{document}